# MULTIPEAKONS OF A TWO-COMPONENT MODIFIED CAMASSA–HOLM EQUATION AND THE RELATION WITH THE FINITE KAC–VAN MOERBEKE LATTICE

XIANG-KE CHANG, XING-BIAO HU, AND JACEK SZMIGIELSKI

ABSTRACT. A spectral and the inverse spectral problem are studied for the two-component modified Camassa–Holm type for measures associated to interlacing peaks. It is shown that the spectral problem is equivalent to an inhomogenous string problem with Dirichlet/Neumann boundary conditions. The inverse problem is solved by Stieltjes's continued fraction expansion, leading to an explicit construction of peakon solutions. Sufficient conditions for global existence of solutions are given. The large time asymptotics reveals that, asymptotically, peakons pair up to form bound states moving with constant speeds. The peakon flow is shown to project to one of the isospectral flows of the finite Kac–van Moerbeke lattice.

## Contents

1991 *Mathematics Subject Classification.* Primary 37K15, 37K40; Secondary 35Q51, 34A55.

The first author was supported in part by the Natural Sciences and Engineering Research Council of Canada (NSERC), the Department of Mathematics and Statistics of the University of Saskatchewan, PIMS postdoctoral fellowship and the Institute of Computational Mathematics, AMSS, CAS.

The second author was supported in part by the National Natural Science Foundation of China #11331008, 11371251; the Knowledge Innovation Program of LSEC, the Institute of Computational Mathematics, AMSS, CAS.

The third author was supported in part by NSERC #163953.

*To appear in Advances in Mathematics.*

## 1. Introduction

One of the most studied integrable equations of the recent two decades has been the Camassa–Holm equation (CH) [6]:

$$m_t + m_x u + 2m u_x = 0, \qquad m = u - u_{xx}, \qquad (x,t) \in \mathbf{R}^2. \tag{1.1}$$

Although it is next to impossible to review even the most relevant developments regarding this equation in a short introduction we nevertheless list the properties of this equation most pertinent to this paper:

(1) the equation is Lax integrable, i.e., it possesses a Lax pair, not only in the smooth sector, but also for a properly restricted class of measures $m$;
(2) it leads to a meaningful model of wave-breaking [13];
(3) the main features of the solutions are captured surprisingly accurately by a class of solutions (multipeakons) coming from finite, discrete measures $m$, for which CH admits an equivalent description in terms of a system of ordinary differential equations and informs a particle interpretation of this class of solutions [6];
(4) the system of ordinary differential equations describing multipeakons can be solved using classical tools of analysis involving continued fractions and orthogonal polynomials [1];
(5) the origin of wave-breaking can be traced back to a simple mechanical phenomenon of a collision of particles [32, 2].

The other features of CH that have attracted a lot of attention are its geometric interpretation as an Euler equation on the Bott–Virasoro group [33, 26] and the existence of accurate numerical integrators capable of handling wave-breaking [12, 21, 9]. Not surprisingly, the initial success of CH also prompted researchers to look for other equations exhibiting similar properties, resulting in several new equations that have been proposed and studied over last twenty years. Here, perhaps the most intriguing is the Degasperis–Procesi equation (DP) [14, 11]

$$m_t + m_x u + 3m u_x = 0, \qquad m = u - u_{xx}, \tag{1.2}$$

which not only supports peakon solutions [30] and wave-breaking [15], but also shocks [29, 42]. Other equations that attracted considerable attention are

(1) the Novikov equation [23, 22, 36]

$$m_t + (m_x u + 3m u_x) u = 0, \qquad m = u - u_{xx}, \tag{1.3}$$

(2) the Hunter–Saxton equation [3, 24]

$$m_t + m_x u + 2m u_x = 0, \qquad m = u_{xx}, \tag{1.4}$$

(3) multicomponent generalizations of CH [8, 18, 19, 27], for example the Geng-Xue (two-component) equation:

$$\begin{aligned} m_t + (m_x u + 3m u_x) v &= 0, \\ n_t + (n_x v + 3n v_x) u &= 0, \end{aligned} \tag{1.5}$$

where $m = u - u_{xx}$ and $n = v - v_{xx}$.

Now we would like to focus our attention on the class of generalizations directly relevant to this work.

### 1.1. The modified CH equation.

The nonlinear partial differential equation

$$m_t + [m(u^2 - u_x^2)]_x = 0, \tag{1.6}$$

where $m = u - u_{xx}$, appeared originally as a new integrable system in the works of Fokas [16] and Fuchssteiner [17] as well as Olver and Rosenau [37]. In the latter, this equation was derived from the general method of tri-Hamiltonian duality applied to the bi-Hamiltonian representation of the modified Korteweg–de Vries equation. We recall that the CH equation can be obtained from the Korteweg–de Vries equation by tri-Hamiltonian duality and hence it is natural to refer to (1.6) as the modified CH equation (mCH), in full agreement with other authors [20, 28]. In [38, 39], the equation was rederived by Qiao and proved by him to possess a Lax pair, a bi-Hamiltonian structure, and new types of soliton solutions.

The subject of our paper is a two-component integrable generalization

$$m_t + [(u - u_x)(v + v_x)m]_x = 0,$$
$$n_t + [(u - u_x)(v + v_x)n]_x = 0,$$
$$m = u - u_{xx}, \qquad n = v - v_{xx},$$

Section 2 gives more details about this equation, in particular its peakon sector. Section 3 is devoted to the (forward) spectral problem associated with this system. We solve the inverse problem in Section 4 for the class of interlacing peakons using continued fractions. We then reinterpret the spectral problem as an inhomogeneous string with mixed boundary conditions. The solution to the inverse problem is subsequently used to give explicit solutions to peakon equations in the interlacing case. In particular, Section 5 contains explicit formulas for interlacing peakons in terms of Hankel determinants. Moreover, in Section 5.2, in particular in Theorem 5.7 we give sufficient conditions for the existence of global multipeakon flows.

Finally, in Section 6 we establish a connection between the peakon sector of the above two-component equation with the Kac–van Moerbeke lattice flows.

## 2. A two-component generalization of the modified CH equation

We briefly review the background needed for the present work. In [40], Song, Qu and Qiao proposed a two-component integrable extension of the modified CH equation (1.6):

$$\begin{aligned} m_t + [(u - u_x)(v + v_x)m]_x &= 0, \\ n_t + [(u - u_x)(v + v_x)n]_x &= 0, \\ m = u - u_{xx}, \qquad n &= v - v_{xx}, \end{aligned} \tag{2.1}$$

which, for simplicity, we shall call 2-mCH. This system of equations is known to possess infinitely many conservation laws as well as a Lax formulation. We mention in passing that in [43] Tian and Liu applied tri-Hamiltonian duality to the bi-Hamiltonian representation of the Wadati-Konno-Ichikawa equation, obtaining 2-mCH equation together with its bi-Hamiltonian structure, thus closing the circle of ideas around tri-Hamiltonian duality. It is elementary to verify that equations (2.1) are compatibility conditions resulting from the Lax pair

$$\frac{\partial}{\partial x}\begin{pmatrix} \Psi_1 \\ \Psi_2 \end{pmatrix} = \frac{1}{2}U\begin{pmatrix} \Psi_1 \\ \Psi_2 \end{pmatrix} \tag{2.2}$$

and

$$\frac{\partial}{\partial t}\begin{pmatrix}\Psi_1\\\Psi_2\end{pmatrix}=\frac{1}{2}V\begin{pmatrix}\Psi_1\\\Psi_2\end{pmatrix}, \tag{2.3}$$

where

$$U=\begin{pmatrix}-1 & \lambda m\\-\lambda n & 1\end{pmatrix}, \tag{2.4}$$

$$V=\begin{pmatrix}4\lambda^{-2}+Q & -2\lambda^{-1}(u-u_x)-\lambda m Q\\2\lambda^{-1}(v+v_x)+\lambda n Q & -Q\end{pmatrix}, \tag{2.5}$$

with $Q=(u-u_x)(v+v_x)$. Note that, for convenience, we have chosen here a slightly different Lax pair from the one in [40]. Furthermore, for $v=u$ the system (2.1) formally reduces to the modified CH equation (1.6) whose Lax pair also follows from that in (2.2) and (2.3).

We are interested in constructing the distributional solutions to the system (2.1) for the case when $m,n$ are discrete measures. Then $u$ and $v$ are piecewise smooth and we can easily define the product of a piecewise smooth function $f$ by $m$ or $n$ as $\langle f\rangle m$ and $\langle f\rangle n$, respectively, where $\langle f\rangle$ means the average of $f$, that is: $\langle f\rangle(a)=\frac{1}{2}\{\lim_{x\to a^+}f(x)+\lim_{x\to a^-}f(x)\}$.

The 2-mCH equation also admits $N$-peakon solutions with two components, similar to the Geng–Xue equation (1.5), [31], obtained by postulating that $u$ and $v$ are given by

$$u=\sum_{k=1}^{N}m_k(t)e^{-|x-x_k(t)|}, \tag{2.6}$$

$$v=\sum_{k=1}^{N}n_k(t)e^{-|x-x_k(t)|}, \tag{2.7}$$

with $m_k n_k=0$, $m_k+n_k>0$ for all $k$. Clearly, in this case $m$ and $n$ are discrete measures, albeit with disjoint support:

$$m=2\sum_{k=1}^{N}m_k\delta_{x_k}, \qquad n=2\sum_{k=1}^{N}n_k\delta_{x_k}. \tag{2.8}$$

The latter ensures that equation (2.2) is a well posed distributional problem. Using elementary distribution calculus (see e.g. [2, 22, 30, 31]) we arrive at the following theorem:

**Theorem 2.1.** *Let $m$ and $n$ be discrete measures associated to $u$ and $v$ given by* (2.6) *and* (2.7). *Then the PDEs* (2.1) *hold if and only if $x_k$, $m_k$ and $n_k$ satisfy the ODEs*

$$\dot{m}_k=0, \qquad \dot{n}_k=0, \tag{2.9}$$

$$\dot{x}_k=(u(x_k)-u_x(x_k))(v(x_k)+v_x(x_k)), \tag{2.10}$$

*where $u_x(x_k)=\langle u_x\rangle(x_k)$ and $v_x(x_k)=\langle v_x\rangle(x_k)$.*

**Remark 2.2.** If we assume that the positions at time $t$ are ordered $x_1(t)<x_2(t)<\cdots<x_N(t)$, then the peakon ODEs (2.9)–(2.10) can be simplified (locally at time $t$) to

$$\dot{x}_k=\left(2\sum_{j<k}m_je^{x_j}+m_ke^{x_k}\right)\left(n_ke^{-x_k}+2\sum_{j>k}n_je^{-x_j}\right), \tag{2.11}$$

with constants $m_j$ and $n_j$.

In this paper, we shall apply the inverse spectral method to solve the peakon ODEs (2.9)–(2.10) for the *interlacing* case of $N = 2K$ sites

$$x_1 < x_2 < \cdots < x_{2K}$$

with the measure $m$ supported on the odd-numbered sites $x_{2k-1}$ and the measure $n$ supported on the even-numbered sites $x_{2k}$. More explicitly, we are going to solve the ODEs

$$\dot{x}_{2k-1} = \left(2\sum_{j=1}^{k-1} m_{2j-1} e^{x_{2j-1}} + m_{2k-1} e^{x_{2k-1}}\right)\left(2\sum_{j=k}^{K} n_{2j} e^{-x_{2j}}\right), \tag{2.12a}$$

$$\dot{x}_{2k} = \left(2\sum_{j=1}^{k} m_{2j-1} e^{x_{2j-1}}\right)\left(n_{2k} e^{-x_{2k}} + 2\sum_{j=k+1}^{K} n_{2j} e^{-x_{2j}}\right), \tag{2.12b}$$

with the constants $m_{2k-1}$, $n_{2k}$ for $k = 1, 2, \ldots, K$, and the initial positions ordered as $x_1(0) < x_2(0) < \cdots < x_{2K}(0)$.

We will choose $m_{2k-1}$ and $n_{2k}$ to have the same sign, positive, which corresponds to peakons. (The negative case of antipeakons can be obtained by the transformation $m_k \mapsto -m_k$, $n_k \mapsto -n_k$.) We will subsequently prove that the spectrum of the associated boundary value problem is real and simple. We will also give a rigorous analysis for the Lax pair of the peakon ODEs in Appendix A ensuring applicability of the inverse spectral method and, finally, we will establish a correspondence between the peakon ODEs and the Kac–van Moerbeke lattice.

## 3. FORWARD SPECTRAL PROBLEM

In this section, we will define a spectral problem and analyze the properties of the spectrum.

3.1. **Modified string problem.** Let us begin with the $x$-part of the Lax pair (2.2). Performing a simple gauge transformation

$$\Psi_1(x) = \Phi_1(x) e^{-x/2}, \qquad \Psi_2(x) = \Phi_2(x) e^{x/2}, \tag{3.1}$$

$$m = 2g e^{-x}, \qquad n = 2h e^{x}, \tag{3.2}$$

equation (2.2) may be rewritten as

$$\frac{\partial}{\partial x}\Phi_1 = \lambda g \Phi_2, \qquad \frac{\partial}{\partial x}\Phi_2 = -\lambda h \Phi_1. \tag{3.3}$$

We now define a spectral problem by imposing the boundary conditions

$$\Phi_1(-\infty) = 0 = \Phi_2(+\infty). \tag{3.4}$$

The eigenvalues of this problem are those values of $\lambda$ for which (3.3) has nontrivial solutions under the restriction (3.4).

**Remark 3.1.** The spectral problem used in [2] to solve the peakon problem for the CH equation involved a string problem with Dirichlet boundary conditions. In our case, if either $g$ or $h$ is a constant, the problem under study also becomes a string problem, but with mixed boundary condition. Therefore, we shall call it a *modified string problem*. This point of view is further supported by the form of the solution to the inverse problem (see Section 4).

**Assumption 3.2.** The positive measures $m$ and $n$ are assumed to be:

(1) discrete and finite, given by

$$m = 2\sum_{k=1}^{K} m_{2k-1}\delta_{x_{2k-1}}, \qquad n = 2\sum_{k=1}^{K} n_{2k}\delta_{x_{2k}}, \tag{3.5}$$

(2) supported on interlacing points $x_1 < x_2 < \cdots < x_{2K}$.

Hence the measures $g$ and $h$ are discrete, with the same support as $m$ and $n$, and take the general form

$$\begin{aligned} g &= \sum_{k=1}^{K} g_k \delta_{x_{2k-1}}, & h &= \sum_{k=1}^{K} h_k \delta_{x_{2k}}, \\ g_k &= m_{2k-1} e^{x_{2k-1}}, & h_k &= n_{2k} e^{-x_{2k}}. \end{aligned} \tag{3.6}$$

The ordering assumption remains true on an open interval containing $t = 0$ provided it holds true for $t = 0$. In the discrete case both $\Phi_1(x)$ and $\Phi_2(x)$ are piecewise constant. Moreover, in the interlacing case, $\Phi_1$ is constant in the intervals $x_{2k-1} < x < x_{2k+1}$ while $\Phi_2(x)$ remains constant in $x_{2k} < x < x_{2k+2}$ with jumps in the values of $\Phi_1$ and $\Phi_2$ at the points $x_{2k-1}$ and $x_{2k}$, respectively (see Figure 1).

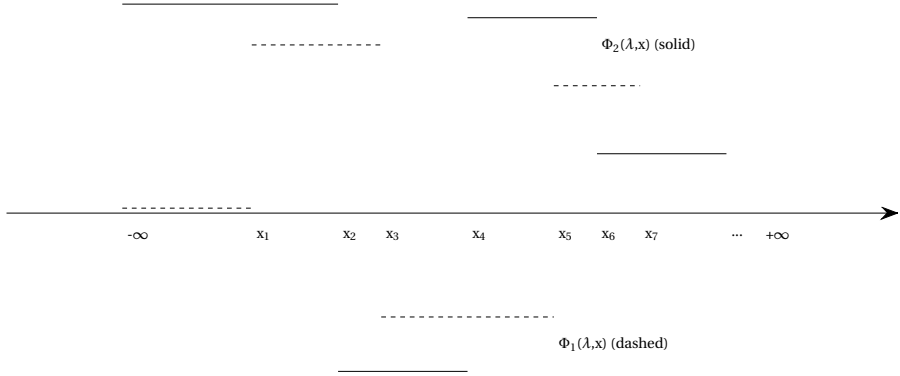

FIGURE 1. Structure of the solution to the interlacing discrete problem (3.3) with initial values $\Phi_1(\lambda, -\infty) = 0$, $\Phi_2(\lambda, -\infty) = 1$, and positive $\lambda$, $g_k$, $h_k$.

Let us define $\Theta_k = \Phi_1(x_{2k+1}-)$ and $\Pi_k = \Phi_2(x_{2k}-)$; then the discrete modified string problem (3.3)–(3.4) becomes

$$\Theta_k - \Theta_{k-1} = \lambda g_k \Pi_k, \qquad 1 \le k \le K, \tag{3.7a}$$

$$\Pi_{k+1} - \Pi_k = -\lambda h_k \Theta_k, \qquad 1 \le k \le K, \tag{3.7b}$$

with boundary conditions: $\quad \Theta_0 = 0, \quad \Pi_{K+1} = 0, \tag{3.7c}$

where $\Pi_{K+1} = \lim_{x\to+\infty} \Phi_2(x)$ and $\Theta_K = \lim_{x\to+\infty} \Phi_1(x)$. Without loss of generality we can impose $\lim_{x\to-\infty} \Phi_2(x) = 1$, which leads to $\Pi_1 = 1$.

The above spectral problem may be written in the matrix form

$$Aq = \lambda Gq, \tag{3.8}$$

where $q = (\Theta_1,\ldots,\Theta_K,\Pi_1,\ldots,\Pi_K)^T$, and

$$A = \begin{pmatrix} 0 & L \\ L^T & 0 \end{pmatrix}, \qquad G = \mathrm{diag}(h_1,\ldots,h_K, g_1,\ldots,g_K),$$

with

$$L = \begin{pmatrix} 1 & -1 & 0 & \cdots & 0 \\ 0 & 1 & -1 & 0 & \\ 0 & 0 & 1 & & \\ \vdots & & & \ddots & -1 \\ 0 & & & 0 & 1 \end{pmatrix}_{K \times K}.$$

Note that $A$ is a symmetric matrix.

Solving the recurrence relations (3.7a)–(3.7b) with the initial conditions $\Theta_0 = 0$ and $\Pi_1 = 1$, we get

$$\Theta_1 = \lambda g_1, \qquad\qquad \Pi_2 = 1 - \lambda^2 g_1 h_1,$$
$$\Theta_2 = \lambda(g_1+g_2) - \lambda^3 g_1 h_1 g_2, \qquad \Pi_3 = 1 - \lambda^2(g_1 h_1 + g_1 h_2 + g_2 h_2) + \lambda^4 g_1 h_1 g_2 h_2,$$

and, more generally, $\Theta_k$ and $\Pi_{k+1}$ are explicitly given by the following formulas:

**Theorem 3.3.** *For any $1 \le k \le K$, we have*

$$\begin{pmatrix} \Theta_k \\ \Pi_{k+1} \end{pmatrix} = \begin{pmatrix} \sum_{l=1}^{k} (-1)^{l+1} \lambda^{2l-1} \sum_{(I,J) \in \widetilde{\Omega}_k^l} g_I h_J \\ 1 + \sum_{l=1}^{k} (-1)^l \lambda^{2l} \sum_{(I,J) \in \Omega_k^l} g_I h_J \end{pmatrix}, \tag{3.9}$$

*where the index sets $\Omega_k^l$ and $\widetilde{\Omega}_k^l$ are defined as*

$$\Omega_k^l = \Big\{ (\{i_1,\ldots,i_l\},\{j_1,\ldots,j_l\}) : 1 \le i_1 \le j_1 < i_2 \le j_2 < \cdots < i_l \le j_l \le k \Big\},$$
$$\widetilde{\Omega}_k^l = \Big\{ (\{i_1,\ldots,i_l\},\{j_1,\ldots,j_{l-1}\}) : 1 \le i_1 \le j_1 < i_2 \le j_2 < \cdots \le j_{l-1} < i_l \le k \Big\},$$

*and $g_I = \prod_{i \in I} g_i$, $h_J = \prod_{j \in J} h_j$, for any index set $I$ (or $J$).*

*Proof.* Recall that the basic recurrence relations (3.7a) and (3.7b) can be written

$$\begin{pmatrix} \Theta_k \\ \Pi_{k+1} \end{pmatrix} = \begin{pmatrix} 1 & 0 \\ -\lambda h_k & 1 \end{pmatrix} \begin{pmatrix} \Theta_k \\ \Pi_k \end{pmatrix}$$
$$= \begin{pmatrix} 1 & 0 \\ -\lambda h_k & 1 \end{pmatrix} \begin{pmatrix} 1 & \lambda g_k \\ 0 & 1 \end{pmatrix} \begin{pmatrix} \Theta_{k-1} \\ \Pi_k \end{pmatrix}$$
$$= \begin{pmatrix} 1 & \lambda g_k \\ -\lambda h_k & -\lambda^2 h_k g_k + 1 \end{pmatrix} \begin{pmatrix} \Theta_{k-1} \\ \Pi_k \end{pmatrix}.$$

The remaining part of the proof proceeds by induction. □

**Corollary 3.4.** *For any $1 \le k \le K$, $\Theta_k$ is an odd polynomial of degree $2k-1$ in $\lambda$ and $\Pi_{k+1}$ is an even polynomial of degree $2k$ in $\lambda$.*

Recall that the spectrum of the problem (3.7a)–(3.7c), or equivalently (3.3)–(3.4), is the set of zeros of $\Pi_{K+1}$. One readily establishes the basic properties of the spectrum.

**Theorem 3.5.** *The spectrum is real and simple, and the eigenvalues appear in pairs of opposite real numbers.*

*Proof.* Noting that $h_k$ and $g_k$ are positive, we let
$$H = \operatorname{diag}(\sqrt{h_1},\ldots,\sqrt{h_K},\sqrt{g_1},\ldots,\sqrt{g_K})$$
so that $G = H^2$. Then (3.8) is equivalent to $H^{-1}AH^{-1}w = \lambda w$, where $w = Hq$. Therefore, the spectrum is real since $H^{-1}AH^{-1}$ is real and symmetric. Moreover, any non-zero solution of (3.3)–(3.4) must be a scalar multiple of a given solution, which implies that there is one eigenfunction for each given $\lambda$ and hence the geometric multiplicity is one. Since $H^{-1}AH^{-1}$ is symmetric, the algebraic multiplicity must be one as well, and thus the spectrum is simple. Finally, if $\lambda$ is a root of $\Pi_{K+1}$, then so is $-\lambda$, since $\Pi_{K+1}$ is an even polynomial. □

3.2. **Time evolution.** For $x > x_{2K}$, we have $\Phi_1(x,\lambda) \equiv \Theta_K(\lambda)$ and $\Phi_2(x,\lambda) \equiv \Pi_{K+1}(\lambda)$. According to the $t$-part (2.3) of the Lax pair, we get

$$\frac{\partial}{\partial t}\Theta_K = \frac{2}{\lambda^2}\Theta_K - \frac{2L}{\lambda}\Pi_{K+1}, \qquad \frac{\partial}{\partial t}\Pi_{K+1} = 0, \tag{3.10}$$

where $L = \sum_{k=1}^{K} g_k$. Thus $\Pi_{K+1}(\lambda)$ is independent of time and its zeros, i.e., the spectrum, are invariant. Consequently, with the help of Theorem 3.3, we obtain explicit formulas for $K$ constants of motion.

**Theorem 3.6.** *The 2-mCH peakon ODEs (2.12a)–(2.12b) admit $K$ constants of motion $(M_1,\ldots,M_K)$, where*
$$M_k = \sum_{(I,J)\in\Omega_K^k} g_I h_J.$$

**Remark 3.7.** The quantity $L$ appearing in the equations of motion above is not a constant of motion (see Theorem 4.3).

**Example 3.8.** As an example of constants of motion we consider the case $K = 3$. The constants of motion, written in the original variables $(m_j, n_j, x_j)$, and with positions $x_j$ satisfying the interlacing assumption 3.2, are

$$\begin{aligned}
M_1 &= m_1 n_2 e^{x_1-x_2} + m_1 n_4 e^{x_1-x_4} + m_1 n_6 e^{x_1-x_6} \\
&\quad + m_3 n_4 e^{x_3-x_4} + m_3 n_6 e^{x_3-x_6} + m_5 n_6 e^{x_5-x_6}, \\
M_2 &= m_1 n_2 m_3 n_4 e^{x_1-x_2+x_3-x_4} + m_1 n_2 m_3 n_6 e^{x_1-x_2+x_3-x_6} \\
&\quad + m_1 n_2 m_5 n_6 e^{x_1-x_2+x_5-x_6} + m_1 n_4 m_5 n_6 e^{x_1-x_4+x_5-x_6} \\
&\quad + m_3 n_4 m_5 n_6 e^{x_3-x_4+x_5-x_6}, \\
M_3 &= m_1 n_2 m_3 n_4 m_5 n_6 e^{x_1-x_2+x_3-x_4+x_5-x_6}.
\end{aligned}$$

## 4. INVERSE SPECTRAL PROBLEM

We first introduce an appropriate Weyl function and proceed to solve the inverse problem of recovering $g_k$ and $h_k$ in terms of the *spectral data*: $\{(\lambda_j^2, b_j) : 1 \le j \le K\}$. Our solution to the inverse problem relies on the use of a theorem of Stieltjes and continued fractions (see [1]).

4.1. **Weyl function.** Let us introduce the Weyl function as

$$W(\lambda) \stackrel{\text{def}}{=} \frac{\Phi_1(\infty, \lambda)}{\lambda \Phi_2(\infty, \lambda)} = \frac{\Theta_K(\lambda)}{\lambda \Pi_{K+1}(\lambda)}. \tag{4.1}$$

Guided by the results from [1] we obtain:

**Theorem 4.1.** *The Weyl function $W$ has the continued fraction expansion*

$$W(\lambda) = \cfrac{1}{-\lambda^2 h_K + \cfrac{1}{g_K + \cfrac{1}{-\lambda^2 h_{K-1} + \cdots + \cfrac{1}{-\lambda^2 h_1 + \cfrac{1}{g_1}}}}}. \tag{4.2}$$

*Proof.* From (3.7a)–(3.7a), we have

$$\frac{\Theta_k}{\Pi_{k+1}} = \frac{\Theta_k}{\Pi_k - \lambda h_k \Theta_k} = \frac{1}{-\lambda h_k + \frac{\Pi_k}{\Theta_k}} = \frac{1}{-\lambda h_k + \frac{\Pi_k}{\Theta_{k-1} + \lambda g_k \Pi_k}} = \frac{1}{-\lambda h_k + \frac{1}{\lambda g_k + \frac{\Theta_{k-1}}{\Pi_k}}}$$

for $k = 1, \ldots, K$, which when applied recursively implies the claim, in view of $\Theta_0 = 0$. □

Additionally, by Theorem 3.5 we can represent the characteristic polynomial of our boundary value problem as

$$\Pi_{K+1}(\lambda) = \prod_{j=1}^{K} (1 - \frac{\lambda^2}{\lambda_j^2})$$

with $0 < \lambda_1^2 < \cdots < \lambda_K^2$, and consequently $W(\lambda)$ can be written in a partial fraction form:

$$W(\lambda) = -\sum_{j=1}^{K} \frac{b_j}{\lambda^2 - \lambda_j^2}, \tag{4.3}$$

which leads to the following theorem.

**Theorem 4.2.** *Let $z = -\lambda^2$, $\zeta_j = \lambda_j^2$ for $1 \leq j \leq K$, and let $d\mu$ denote the discrete measure $d\mu = \sum_{j=1}^{K} b_j \delta_{\zeta_j}$. Then the Weyl function is the Stieltjes transform of the measure $d\mu$, that is*

$$W(\lambda) = \int \frac{d\mu(\zeta)}{z + \zeta}. \tag{4.4}$$

Now we turn our attention to the study of $\{b_j\}$.

**Theorem 4.3.** *The residues $\{b_j\}$ are always positive and they evolve according to*

$$\dot{b}_j = \frac{2b_j}{\lambda_j^2}. \tag{4.5}$$

*The quantity $L = \sum_{k=1}^{K} g_k$ appearing in the evolution equations* (3.10) *can be computed from $L = \sum_{j=1}^{K} \frac{b_j}{\lambda_j^2}$.*

*Proof.* From (3.10) we obtain

$$\dot{W}(\lambda) = \frac{2W(\lambda)}{\lambda^2} - \frac{2L}{\lambda^2},$$

which together with (4.3) implies $L = W(\lambda)|_{\lambda=0} = \sum_{j=1}^{K} \frac{b_j}{\lambda_j^2}$. Substituting the expression (4.3) into the above relation and comparing the residues on both sides, we obtain the evolution equation (4.5).

As for the positivity of $b_j$ it is sufficient to prove $b_j(0) > 0$, since the evolution relation (4.5) leads to $b_j(t) = b_j(0)e^{2t/\lambda_j^2}$. The positivity of $b_j(0)$ will be proved in Corollary 6.5. $\square$

4.2. **Stieltjes's theorem.** A classical result of Stieltjes [1, 2, 41] allows one to solve the inverse spectral problem corresponding to peakon ODEs of the CH equation [1, 2]. This theorem also applies to our case and, as a consequence, $g_k$ and $h_k$ can be recovered from the relatively simple formulas involving $b_j$ and $\lambda_j$. For the reader's convenience we recall now Stieltjes's theorem.

**Theorem 4.4** (T. Stieltjes). *The Laurent series*

$$V(z) = \sum_{k=0}^{\infty} \frac{(-1)^k A_k}{z^{k+1}},$$

*whose coefficients $A_k$ are strictly positive, can be uniquely written as an infinite continued fraction*

$$V(z) = \cfrac{1}{za_1 + \cfrac{1}{a_2 + \cfrac{1}{za_3 + \cdots}}}, \qquad a_i > 0,$$

*if the Hankel determinants $H_k^l = \det(A_{i+j+l})_{i,j=0}^{k-1}$ satisfy*

$$H_k^0 > 0, \qquad H_k^1 > 0, \qquad \text{for all } k > 0. \tag{4.6}$$

*Then*

$$a_{2k} = \frac{(H_k^0)^2}{H_k^1 H_{k-1}^1}, \qquad a_{2k+1} = \frac{(H_k^1)^2}{H_k^0 H_{k+1}^0},$$

*with the proviso that $H_0^0 = H_0^1 = 1$.*

*If $V(z)$ is a rational function then the condition* (4.6) *is replaced with*

$$H_k^0 > 0, \qquad H_j^1 > 0, \qquad k \leq K, \qquad j \leq J \qquad \text{for some } K \text{ and } J.$$

From Stieltjes's theorem as well as our earlier Theorems 4.1 and 4.2, we can easily recover $g_k$ and $h_k$.

**Theorem 4.5.** *For $1 \leq k \leq K$, the coefficients $g_k$ and $h_k$ admit the explicit formulas in terms of $\lambda_j$ and $b_j$:*

$$g_{K+1-k} = \frac{(H_k^0)^2}{H_k^1 H_{k-1}^1}, \qquad h_{K+1-k} = \frac{(H_{k-1}^1)^2}{H_k^0 H_{k-1}^0}, \tag{4.7}$$

*where the Hankel determinant $H_k^l$ is given by $H_k^l = \det(A_{i+j+l})_{i,j=0}^{k-1}$ for $l \geq 0$, with moments $A_k \stackrel{\text{def}}{=} \int \zeta^k d\mu(\zeta) = \sum_{j=1}^{K} \lambda_j^{2k} b_j$.*

**Remark 4.6.** Only $H_k^0$ and $H_k^1$ appear in the denominators of the explicit expressions for $g_k$ and $h_k$. In Section 5 we show for the sake of self-completeness that both these determinants are strictly positive.

4.3. **Modified string problem versus string problem; a comparison.** Let us consider an inhomogeneous string of length $L$:

$$-v'' = \zeta \rho v, \quad v(0) = 0, \quad v'(L) = 0, \tag{4.8}$$

with the Dirichlet boundary condition on the left end of the string and the Neumann boundary condition on the right end. Suppose that the mass density $\rho$ is a finite discrete measure $\rho = \sum_{k=1}^K \rho_k \delta_{\xi_k}$, associated with a collection of point-masses $\rho_k > 0$ placed at $\xi_k$. In every open interval $\xi_k < \xi < \xi_{k+1}$ we have

$$v(\xi) = p_{k+1}(\xi - \xi_k) + v_k,$$

and the function $v$ remains continuous at the positions of the masses, while the derivative $v'$ has the jump

$$[v'](\xi_k) = -\zeta \rho_k v_k \tag{4.9}$$

there. The distributional equation (4.8) can be equivalently written as a system in $p_k$ and $v_k$:

**Lemma 4.7.** *Let $l_k = \xi_k - \xi_{k-1}$ denote the distance between the masses $\rho_k$ and $\rho_{k-1}$. Then equation (4.8) is equivalent to*

$$v_k - v_{k-1} = l_k p_k, \qquad 1 \leq k \leq K, \tag{4.10a}$$

$$p_{k+1} - p_k = -\zeta \rho_k v_k, \qquad 1 \leq k \leq K, \tag{4.10b}$$

$$v_0 = 0, \quad p_{K+1} = 0, \tag{4.10c}$$

Comparing equations (4.10) with equations (3.7) reveals the essence of the connection between these two boundary value problems. One obtains the following theorem.

**Theorem 4.8.** *For fixed constants $(m_j, n_j)$, let $F$ be the map, specified below, between the discrete modified string problem (3.3)–(3.4) with interlacing data and the string problem (4.8)*

$$F: \begin{cases} g_k \mapsto l_k, \\ h_k \mapsto \rho_k, \\ \lambda \mapsto \zeta = \lambda^2 \\ \Pi_k(\lambda) \mapsto p_k(\zeta) = \lambda \Pi_k(\lambda), \\ \Theta_k(\lambda) \mapsto v_k(\zeta) = \Theta_k(\lambda). \end{cases}$$

*The map $F$ is $2$-to-$1$.*

*Proof.* The map $F$ is well defined because $\Theta_k(\lambda)$ is an even function of $\lambda$ and $\Pi_k(\lambda)$ is odd. $\square$

**Remark 4.9.** Based on the theorem above we see that for fixed constants $(m_j, n_j)$ the boundary value problem (3.3)–(3.4) with interlacing data is the unique *square root* of the string problem (4.8) considered on the Riemann surface of $\sqrt{\zeta}$. It is in this sense that we can view both boundary value problems as equivalent.

For completeness we record the evolution of the spectral measure of the string. By combining Theorem 4.2 and formula (4.5) we obtain

$$d\mu_t = \sum_{j=1}^{K} b_j(0) e^{\frac{2t}{\zeta_j}} \delta_{\zeta_j} = e^{\frac{2t}{\zeta}} d\mu_0. \tag{4.11}$$

Let us conclude this subsection by summarizing how the quantities of interests, namely $g_k$ and $h_k$, behave under the correspondence $F$:

$$g_k \xrightarrow{F} \text{distances } l_k,$$

$$h_k \xrightarrow{F} \text{masses } \rho_k.$$

Under this correspondence, the minimal length of the string corresponding to the discrete modified string is precisely $L = \sum_{k=1}^{K} l_k = \sum_{k=1}^{K} g_k$ (see equation (3.10)). In particular, this means that under the time-flow given by the 2-mCH equation **the length of the string varies**, in contrast to the CH case [1].

## 5. MULTIPEAKONS

5.1. **Closed form of multipeakons.** In this subsection we give explicit formulas for multipeakon solutions to the 2-mCH equation (2.1) in the interlacing case. Combining (3.6) and Theorem 4.5, we obtain the following theorem:

**Theorem 5.1.** *The 2-mCH equation* (2.1) *admits the multipeakon solution*

$$\begin{aligned} u(x,t) &= \sum_{j=1}^{K} m_{2j-1} \exp(-|x - x_{2j-1}(t)|), \\ v(x,t) &= \sum_{j=1}^{K} n_{2j} \exp(-|x - x_{2j}(t)|), \end{aligned} \tag{5.1}$$

*with*

$$\begin{aligned} x_{2K+1-2k} &= \ln\left(\frac{1}{m_{2K+1-2k}} \cdot \frac{(H_k^0)^2}{H_k^1 H_{k-1}^1}\right), \\ x_{2K+2-2k} &= \ln\left(n_{2K+2-2k} \cdot \frac{H_k^0 H_{k-1}^0}{(H_{k-1}^1)^2}\right), \end{aligned} \tag{5.2}$$

*and the positive constants $m_{2j-1}$ and $n_{2j}$. Here, $H_k^l(t) = \det(A_{i+j+l}(t))_{i,j=0}^{k-1}$, and the moments $A_k(t)$ are given by*

$$A_k(t) = \int \zeta^k d\mu_t(\zeta),$$

*with $d\mu_t$ defined by* (4.11).

For applications it is important that the Hankel determinants $H_k^l(t)$ can be evaluated explicitly. To this end we prove the following lemma.

**Lemma 5.2.** *Let $\lambda_j^2 = \zeta_j$ and $l \geq 0$. Then the Hankel determinants can be computed as follows:*

$$H_k^l = \sum_{J \in \binom{[1,K]}{k}} b_J \zeta_J^l \Delta_J^2, \qquad 0 \leq k \leq K, \tag{5.3}$$

where $\binom{[1,K]}{k}$ denotes the set of $k$-element subsets $J = \{j_1 < \cdots < j_k\}$ of the integer interval $[1, K] = \{1, \ldots, K\}$, i.e.,

$$\binom{[1,K]}{k} = \{J = \{j_1, \ldots, j_k\} : 1 \leq j_1 < \cdots < j_k \leq K\},$$

and

$$b_J = \prod_{j \in J} b_j, \qquad \zeta_J^l = \prod_{j \in J} \zeta_j^l, \qquad \Delta_J^2 = \prod_{i,j \in J, i<j} (\zeta_j - \zeta_i)^2.$$

In particular,

$$H_0^l = 1$$

and

$$H_K^l = b_{[1,K]} \zeta_{[1,K]}^l \Delta_{[1,K]}^2.$$

Additionally, for $l \geq 0$, we have

$$H_k^l = 0, \qquad k > K.$$

**Remark 5.3.** The positivity of $H_k^l$ for $l \geq 0$, $0 \leq k \leq K$ immediately follows from this lemma. Similar formulas have appeared in [2, 7].

*Proof.* The result for $k = 0$ holds by convention. The case $k > 0$ follows from Heine's formula

$$H_k^l = \frac{1}{k!} \int \zeta_1^l \ldots \zeta_k^l \prod_{1 \leq i < j \leq k} (\zeta_j - \zeta_i)^2 d\mu(\zeta_1) \ldots d\mu(\zeta_k),$$

applied to the measure $d\mu = \sum_{j=1}^{K} b_j \delta_{\zeta_j}$, and the identification $\zeta_j = \lambda_j^2$.

The determinant $H_k^l$ vanishes for $k > K$, since the rank of the matrix $(A_{i+j+l})_{i,j=0}^{k-1}$ is not greater than $K$. □

For the sake of illustration of our results we give now explicit formulas for peakon solutions for the cases $K = 1, 2, 3$. The notation is that of Lemma 5.2. We assume that at the initial time the positions satisfy the interlacing condition which subsequently will hold in an open time interval around the initial time; in this sense the solutions below are only local solutions. The very relevant question of the existence of solutions that are global in $t$ will be taken up in section 5.2.

**Example 5.4.** The $1 + 1$ peakon solution is

$$x_1 = \ln\left(\frac{1}{m_1} \cdot \frac{b_1}{\zeta_1}\right), \qquad x_2 = \ln(n_2 \cdot b_1).$$

**Example 5.5.** The $2 + 2$ multipeakon solution is

$$x_1 = \ln\left(\frac{1}{m_1} \cdot \frac{b_1 b_2 (\zeta_2 - \zeta_1)^2}{\zeta_1 \zeta_2 (b_1 \zeta_1 + b_2 \zeta_2)}\right),$$

$$x_2 = \ln\left(n_2 \cdot \frac{b_1 b_2 (b_1 + b_2)(\zeta_2 - \zeta_1)^2}{(b_1 \zeta_1 + b_2 \zeta_2)^2}\right),$$

$$x_3 = \ln\left(\frac{1}{m_3} \cdot \frac{(b_1 + b_2)^2}{b_1 \zeta_1 + b_2 \zeta_2}\right),$$

$$x_4 = \ln(n_4 \cdot (b_1 + b_2)).$$

**Example 5.6.** The $3+3$ multipeakon solution is

$$x_1 = \ln\left(\frac{1}{m_1} \cdot \frac{b_1 b_2 b_3 \Delta_{123}}{\zeta_1\zeta_2\zeta_3\,(b_1 b_2 \zeta_1 \zeta_2 \Delta_{12} + b_2 b_3 \zeta_2 \zeta_3 \Delta_{23} + b_1 b_3 \zeta_1 \zeta_3 \Delta_{13})}\right),$$

$$x_2 = \ln\left(n_2 \cdot \frac{b_1 b_2 b_3 \Delta_{123}\,(b_1 b_2 \Delta_{12} + b_2 b_3 \Delta_{23} + b_1 b_3 \Delta_{13})}{(b_1 b_2 \zeta_1 \zeta_2 \Delta_{12} + b_2 b_3 \zeta_2 \zeta_3 \Delta_{23} + b_1 b_3 \zeta_1 \zeta_3 \Delta_{13})^2}\right),$$

$$x_3 = \ln\Biggl(\frac{1}{m_3} \cdot \frac{(b_1 b_2 \Delta_{12} + b_2 b_3 \Delta_{23} + b_1 b_3 \Delta_{13})^2}{(b_1 \zeta_1 + b_2 \zeta_2 + b_3 \zeta_3)}$$
$$\cdot \frac{1}{(b_1 b_2 \zeta_1 \zeta_2 \Delta_{12} + b_2 b_3 \zeta_2 \zeta_3 \Delta_{23} + b_1 b_3 \zeta_1 \zeta_3 \Delta_{13})}\Biggr),$$

$$x_4 = \ln\left(n_4 \cdot \frac{(b_1 + b_2 + b_3)\,(b_1 b_2 \Delta_{12} + b_2 b_3 \Delta_{23} + b_1 b_3 \Delta_{13})}{(b_1 \zeta_1 + b_2 \zeta_2 + b_3 \zeta_3)^2}\right),$$

$$x_5 = \ln\left(\frac{1}{m_5} \cdot \frac{(b_1 + b_2 + b_3)^2}{b_1\zeta_1 + b_2\zeta_2 + b_3\zeta_3}\right),$$

$$x_6 = \ln\left(n_6 \cdot (b_1 + b_2 + b_3)\right).$$

5.2. **Global existence of multipeakon flows.** In this subsection we will formulate sufficient conditions under which the multipeakon flows exist globally. First, we will extend our multi-index notation: given two disjoint sets $A \subset [1, K]$ and $B \subset [1, K]$ we will denote by $\Delta^2_{A,B}$ the square of the generalized Vandermonde determinant

$$\Delta^2_{A,B} = \prod_{i \in A,\ j \in B} (\zeta_i - \zeta_j)^2.$$

**Theorem 5.7** (Global existence of multipeakons)**.** *Given arbitrary spectral data*

$$\{b_j > 0, \zeta_j > 0 : 1 \le j \le K, \zeta_j < \zeta_{j+1}\},$$

*suppose the masses $m_{2k-1}$ and $n_{2k}$ satisfy*

$$M_k < m_{2k-1} n_{2k}, \qquad 1 \le k \le K, \tag{5.4a}$$

$$N_k > n_{2k} m_{2k+1}, \qquad 1 \le k \le K-1, \tag{5.4b}$$

*where*

$$M_k = \frac{\zeta_K^{K-k}}{\zeta_1^{K+1-k}}, \qquad N_k = \frac{\zeta_1^{K-k}}{(K-k)\zeta_K^{K-1-k}} \frac{\left(\min_i(\zeta_{i+1} - \zeta_i)\right)^{2(K-1-k)}}{\left(\zeta_K - \zeta_1\right)^{2(K-k)}}. \tag{5.5}$$

*Then the multipeakon solutions* (5.1) *exist for all $t \in \mathbf{R}$.*

*Proof.* The solutions described in equation (5.2) are valid multipeakon solutions as long as the interlacing conditions hold. We write these conditions as:

$$x_{2k'-1} < x_{2k'}, \qquad 1 \le k' \le K,$$
$$x_{2k'} < x_{2k'+1}, \qquad 1 \le k' \le K-1,$$

where $k' = K - k + 1$, and use equations (5.2) to express these inequalities in terms of Hankel determinants, resulting in

$$\frac{H_k^0 H_{k-1}^1}{H_{k-1}^0 H_k^1} < m_{2K+1-2k} n_{2K+2-2k}, \qquad 1 \le k \le K, \tag{5.6a}$$

$$\frac{H_{k-1}^0 H_{k-1}^1}{H_k^0 H_{k-2}^1} > n_{2K+2-2k} m_{2K+1-2(k-1)}, \qquad 2 \le k \le K. \tag{5.6b}$$

Below, we establish the following uniform bounds (with respect to $t$):

$$\frac{H_k^0 H_{k-1}^1}{H_{k-1}^0 H_k^1} < \frac{\zeta_K^{k-1}}{\zeta_1^k}, \qquad 1 \le k \le K, \qquad (5.7a)$$

$$\frac{H_{k-1}^0 H_{k-1}^1}{H_k^0 H_{k-2}^1} > \frac{\zeta_1^{k-1}}{(k-1)\zeta_K^{k-2}} \frac{(\min(\zeta_{i+1} - \zeta_i))^{2(k-2)}}{(\zeta_K - \zeta_1)^{2(k-1)}}, \qquad 2 \le k \le K. \qquad (5.7b)$$

To conclude the proof of the theorem we then just need to shift $k \mapsto K + 1 - k$ and use these bounds to define the constants $M_k$ and $N_k$. $\square$

*Proof of* (5.7a). Since $H_j^l = \sum_{J \in \binom{[1,K]}{j}} b_J \zeta_J^l \Delta_J^2$ we get the upper bound

$$H_j^1 < (\zeta_K)^j H_j^0 \qquad (5.8)$$

and the lower bound

$$H_j^1 > (\zeta_1)^j H_j^0, \qquad (5.9)$$

which implies the claim. $\square$

*Proof of* (5.7b). Equations (5.8)–(5.9) imply

$$\frac{H_{k-1}^0 H_{k-1}^1}{H_k^0 H_{k-2}^1} > \frac{\zeta_1^{k-1}}{\zeta_K^{k-2}} \frac{H_{k-1}^0 H_{k-1}^0}{H_k^0 H_{k-2}^0}.$$

We recall that by the definition of Hankel determinants,

$$(H_{k-1}^0)^2 = \sum_{A,B \in \binom{[1,K]}{k-1}} b_A b_B \Delta_A^2 \Delta_B^2.$$

Now we turn to the denominator $H_k^0 H_{k-2}^0$ which can be written

$$H_k^0 H_{k-2}^0 = \sum_{\substack{I \in \binom{[1,K]}{k-2} \\ J \in \binom{[1,K]}{k}}} b_I b_J \Delta_I^2 \Delta_J^2.$$

Since the cardinality of $J$ exceeds that of $I$ it is always possible to find a unique smallest index $i$ in $J$ which is not in $I$. This leads to the map

$$\Phi : \binom{[1,K]}{k-2} \times \binom{[1,K]}{k} \longrightarrow \binom{[1,K]}{k-1} \times \binom{[1,K]}{k-1}, \qquad (5.10)$$

$$(I, J) \longmapsto (I \cup \{i\}, J \setminus \{i\}),$$

where $2 \le k \le K$. Let us now define $A = I \cup \{i\}$, $B = J \setminus \{i\}$. Clearly $b_I b_J = b_A b_B$ and $A, B \in \binom{[1,K]}{k-1}$. Moreover,

$$\Delta_I^2 \Delta_J^2 = \Delta_A^2 \Delta_B^2 \frac{\Delta_{\{i\},B}^2}{\Delta_{\{i\},I}^2} < \frac{(\zeta_K - \zeta_1)^{2(k-1)}}{\min_j(\zeta_{j+1} - \zeta_j)^{2(k-2)}} \Delta_A^2 \Delta_B^2$$

which implies

$$H_k^0 H_{k-2}^0 < \frac{(\zeta_K - \zeta_1)^{2(k-1)}}{\min_j(\zeta_{j+1} - \zeta_j)^{2(k-2)}} \sum_{\substack{A,B \in \binom{[1,K]}{k-1} \\ (A,B) \in \text{Image}(\Phi)}} \#[\Phi^{-1}(A,B)] b_A b_B \Delta_A^2 \Delta_B^2,$$

where $\#[\Phi^{-1}(A, B)]$ counts the number of pairs $(I, J)$ which are mapped by $\Phi$ into the same $(A, B)$. First, by construction, $\#[\Phi^{-1}(A, B)] \le 1$ for $k = 2$. When $k \ge 3$ then two distinct pairs $(I_1, J_1) \neq (I_2, J_2)$ are mapped to the same $(A, B)$ if, for the smallest $i_1 \in J_1 \setminus I_1$ and the smallest $i_2 \in J_2 \setminus I_2$, there exists $L \in \binom{[1,K]}{k-3}, M \in \binom{[1,K]}{k-1}$ such that

$$I_1 = L \cup \{i_2\}, \qquad J_1 = M \cup \{i_1\},$$
$$I_2 = L \cup \{i_1\}, \qquad J_2 = M \cup \{i_2\},$$

in which case $A = L \cup \{i_1\} \cup \{i_2\}$, $B = M$. Thus $\#[\Phi^{-1}(A, B)]$ is bounded from above by the number of ways we can select an individual entry from $A$, hence $\#[\Phi^{-1}(A, B)] \leq k-1$, and

$$\begin{aligned} H_k^0 H_{k-2}^0 &< \frac{(k-1)(\zeta_K - \zeta_1)^{2(k-1)}}{\min_j (\zeta_{j+1} - \zeta_j)^{2(k-2)}} \sum_{\substack{A,B \in \binom{[1,K]}{k-1} \\ (A,B) \in \text{Image}(\Phi)}} b_A b_B \Delta_A^2 \Delta_B^2 \\ &< \frac{(k-1)(\zeta_K - \zeta_1)^{2(k-1)}}{\min_j (\zeta_{j+1} - \zeta_j)^{2(k-2)}} \sum_{A,B \in \binom{[1,K]}{k-1}} b_A b_B \Delta_A^2 \Delta_B^2 \\ &= \frac{(k-1)(\zeta_K - \zeta_1)^{2(k-1)}}{\min_j (\zeta_{j+1} - \zeta_j)^{2(k-2)}} (H_{k-1}^0)^2, \end{aligned} \tag{5.11}$$

which, in turn, proves the bound

$$\frac{H_{k-1}^0 H_{k-1}^1}{H_k^0 H_{k-2}^1} > \frac{\zeta_1^{k-1}}{(k-1)\zeta_K^{k-2}} \frac{(\min(\zeta_{i+1} - \zeta_i))^{2(k-2)}}{(\zeta_K - \zeta_1)^{2(k-1)}}, \qquad 3 \leq k \leq K.$$

Finally, when $k = 2$, one easily obtains

$$\frac{H_1^0 H_1^1}{H_2^0} > \frac{2\zeta_1}{(\zeta_K - \zeta_1)^2} > \frac{\zeta_1}{(\zeta_K - \zeta_1)^2},$$

hence (5.7b) holds for $2 \leq k \leq K$. $\square$

To illustrate the global existence we consider an example with $K = 2$.

**Example 5.8.** Let $K = 2$ and $b_1(0) = 1$, $b_2(0) = 2$, $\zeta_1 = 3$, $\zeta_2 = 6$, $m_1 = 2$, $n_2 = 1$, $m_3 = 0.3$, $n_4 = 2$. It is easy to show that the conditions in Theorem 5.7 are satisfied, i.e.,

$$m_1 n_2 > \frac{\zeta_2}{\zeta_1^2}, \qquad m_3 n_4 > \frac{1}{\zeta_1}, \qquad n_2 m_3 < \frac{\zeta_1}{(\zeta_2 - \zeta_1)^2}.$$

Hence the order $x_1 < x_2 < x_3 < x_4$ will be preserved at all time and one can use the explicit formulas for the 2-peakon solution at all time, resulting in the graphs shown in Figure 2.

**Remark 5.9.** We derived earlier a general sufficient condition for the global existence for any $K$. However, for small $K$, one can construct other sufficient conditions by exploring other types of inequalities available for small $K$. For example, one sufficient condition for $K = 2$ is

$$m_1 n_2 \geq 1, \qquad m_3 n_4 \geq 1, \qquad n_2 m_3 \leq 1,$$
$$\zeta_2 > \zeta_1 \geq 1, \qquad \zeta_1 + \zeta_2 \geq (\zeta_2 - \zeta_1)^2,$$

which ensures that $x_1 < x_2 < x_3 < x_4$ for any time.

**Example 5.10.** We choose $b_1(0) = 1$, $b_2(0) = 2$, $\zeta_1 = 3$, $\zeta_2 = 6$, $m_1 = 2$, $n_2 = 1$, $m_3 = 0.9$, $n_4 = 1.2$ so that

$$m_1 n_2 \geq 1, \qquad m_3 n_4 \geq 1, \qquad n_2 m_3 \leq 1,$$
$$\zeta_2 > \zeta_1 \geq 1, \qquad \zeta_1 + \zeta_2 \geq (\zeta_2 - \zeta_1)^2.$$

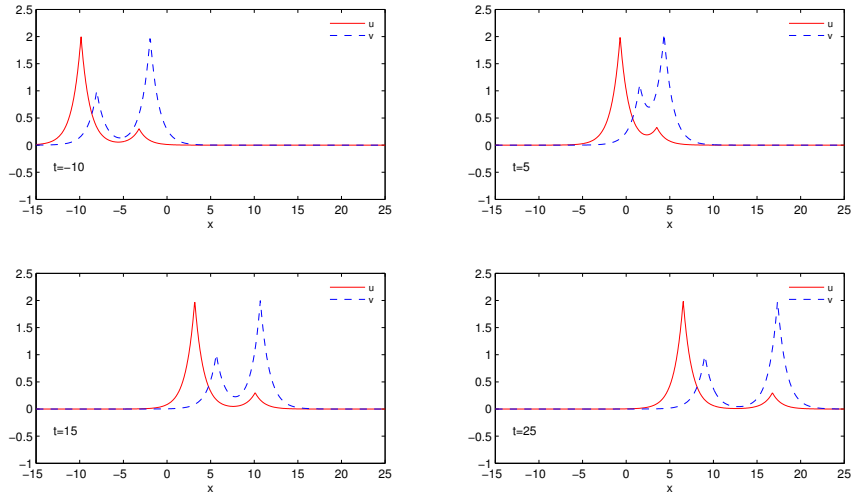

FIGURE 2. A 2 + 2-peakon solution at times $t = -10, 5, 15, 25$ with $b_1(0) = 1$, $b_2(0) = 2$, $\zeta_1 = 3$, $\zeta_2 = 6$, $m_1 = 2$, $n_2 = 1$, $m_3 = 0.3$, $n_4 = 2$.

Graphs illustrating this case are presented in Figure 3. We emphasize that the parameters $m_1$, $n_2$, $m_3$, $n_4$ do not satisfy the condition in Theorem 5.7.

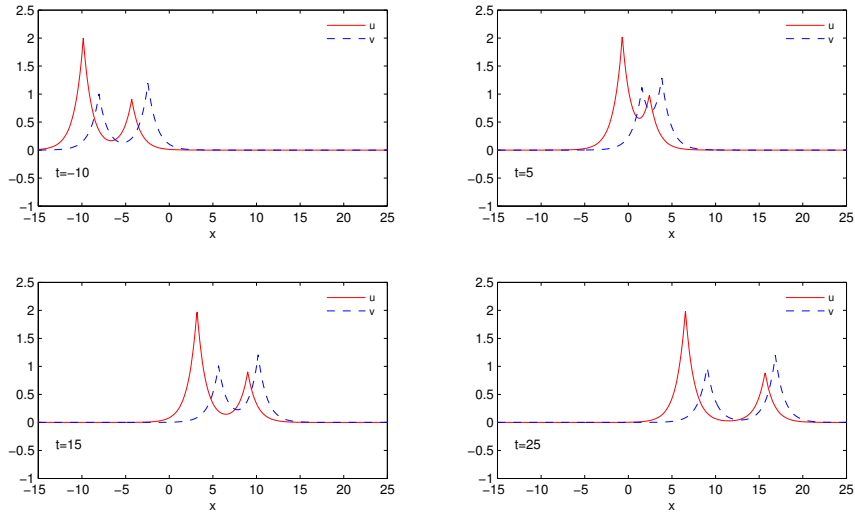

FIGURE 3. A 2 + 2-peakon solution at time $t = -10, 5, 15, 25$ corresponding to the choice $b_1(0) = 1$, $b_2(0) = 2$, $\zeta_1 = 3$, $\zeta_2 = 6$, $m_1 = 2$, $n_2 = 1$, $m_3 = 0.9$, $n_4 = 1.2$.

5.3. **Asymptotics.** Once the determinants $H_k^0$ and $H_k^1$ are known, as well as assuming $\zeta_j$ to be ordered so that $0 < \zeta_1 \cdots < \zeta_K$, we may give the long-term asymptotics for multipeakons by noting that the dominant term corresponds to either

$$\prod_{j=1}^{k} b_j(t) = \prod_{j=1}^{k} b_j(0) \exp\left(\frac{2t}{\zeta_j}\right), \qquad \text{as } t \to +\infty$$

or

$$\prod_{j=k}^{K} b_j(t) = \prod_{j=k}^{K} b_j(0) \exp\left(\frac{2t}{\zeta_j}\right), \qquad \text{as } t \to -\infty.$$

By employing Theorem 5.1 and Lemma 5.2, we then easily obtain the following theorem:

**Theorem 5.11.** *Suppose that the masses $m_j$ and $n_j$ satisfy the conditions of Theorem 5.7. Then, as $t \to +\infty$, for $l \geq 0$ and $1 \leq k \leq K$, we have*

$$H_k^l \sim \left(\prod_{j=1}^{k} b_j(0)\zeta_j^l\right)\left(\prod_{1 \leq i < j \leq k} (\zeta_j - \zeta_i)^2\right)\exp\left(\sum_{j=1}^{k} \frac{2t}{\zeta_j}\right),$$

$$x_{2K+1-2k} \sim \frac{2t}{\zeta_k} + \ln\frac{b_k(0)}{m_{2K+1-2k}\zeta_k} + \ln\left(\prod_{j=1}^{k-1}\left(\frac{\zeta_k}{\zeta_j} - 1\right)^2\right),$$

$$x_{2K+2-2k} \sim \frac{2t}{\zeta_k} + \ln(b_k(0)n_{2K+2-2k}) + \ln\left(\prod_{j=1}^{k-1}\left(\frac{\zeta_k}{\zeta_j} - 1\right)^2\right).$$

*Likewise, as $t \to -\infty$, for $l \geq 0$ and $1 \leq k \leq K$, we have*

$$H_{k+1-k}^l \sim \left(\prod_{j=k}^{K} b_j(0)\zeta_j^l\right)\left(\prod_{k \leq i < j \leq K} (\zeta_j - \zeta_i)^2\right)\exp\left(\sum_{j=k}^{K} \frac{2t}{\zeta_j}\right),$$

$$x_{2k-1} \sim \frac{2t}{\zeta_k} + \ln\frac{b_k(0)}{m_{2k-1}\zeta_k} + \ln\left(\prod_{j=k+1}^{K}\left(\frac{\zeta_k}{\zeta_j} - 1\right)^2\right),$$

$$x_{2k} \sim \frac{2t}{\zeta_k} + \ln(b_k(0)n_{2k}) + \ln\left(\prod_{j=k+1}^{K}\left(\frac{\zeta_k}{\zeta_j} - 1\right)^2\right).$$

Recall that the function $u$ is a superposition of peakons centred at $x_{2k-1}$, and $v$ is a superposition of peakons centered at $x_{2k}$. It follows from the above theorem that **asymptotically, as $t \to \infty$, the peaks of $u$ and $v$ arrange themselves in rigid pairs of peaks of type $u$ and $v$ with the asymptotic velocity of the $k$th pair being $\frac{2}{\zeta_{K+1-k}}$. A similar phenomenon occurs as $t \to -\infty$ where the asymptotic velocity is $\frac{2}{\zeta_k}$.**

Moreover, the term with asymptotic velocity $\frac{2}{\zeta_k}$ undergoes a phase shift:

$$\lim_{t \to +\infty}\left(x_{2K+1-2k} - \frac{2t}{\zeta_k}\right) - \lim_{t \to -\infty}\left(x_{2k-1} - \frac{2t}{\zeta_k}\right)$$

$$= \ln\frac{m_{2k-1}}{m_{2K+1-2k}} + \ln\left(\prod_{j=1}^{k-1}\left(\frac{\zeta_k}{\zeta_j} - 1\right)^2\right) - \ln\left(\prod_{j=k+1}^{K}\left(\frac{\zeta_k}{\zeta_j} - 1\right)^2\right)$$

and

$$\lim_{t \to +\infty}\left(x_{2K+2-2k} - \frac{2t}{\zeta_k}\right) - \lim_{t \to -\infty}\left(x_{2k} - \frac{2t}{\zeta_k}\right)$$

$$= \ln\frac{n_{2K+2-2k}}{n_{2k}} + \ln\left(\prod_{j=1}^{k-1}\left(\frac{\zeta_k}{\zeta_j} - 1\right)^2\right) - \ln\left(\prod_{j=k+1}^{K}\left(\frac{\zeta_k}{\zeta_j} - 1\right)^2\right).$$

This is similar to the CH and DP cases of peakons with one important difference: the CH and DP peakons *scatter*, meaning that they become asymptotically separated (the distances between peaks diverge), while the 2-mCH peakons form **bound states of pairs** consisting of adjacent peaks of type $u$ and $v$ and only these pairs scatter. This indicates that the 2-mCH theory exhibits features of an integrable theory of composite systems, a property which has also been observed in the case of the Geng–Xue equation [31].

Finally, to illustrate our point made above, positing that the 2-mCH equations describe a composite system, we provide two graphs of the field $uv$. It is worth stressing that the rectangular shapes emerge naturally from the multiplicative superposition of peaks and from the asymptotic behaviour, not from the choice of initial conditions.

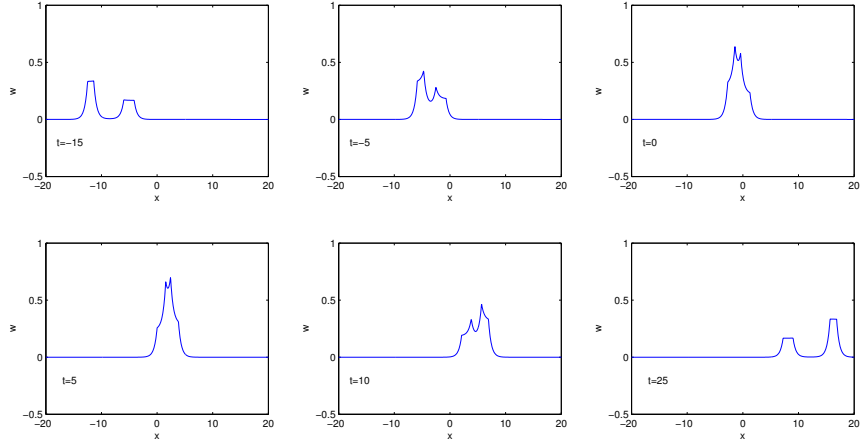

FIGURE 4. Graphs of $w(x,t) = u(x,t)v(x,t)$ for $K = 2$ at $t = -15, -5, 0, 5, 10, 25$ in the case of $b_1(0) = 1$, $b_2(0) = 2$, $\zeta_1 = 3$, $\zeta_2 = 6$, $m_1 = 2$, $n_2 = 1$, $m_3 = 0.9$, $n_4 = 1.2$.

## 6. PEAKONS ODEs AND THE KAC–VAN MOERBEKE LATTICE

In this section, we present a correspondence between the peakon system (2.12a)–(2.12b) and the finite Kac–van Moerbeke lattice which is quite analogous to the relation between the peakon system of the CH equation and the finite Toda lattice [4].

### 6.1. The discrete modified string and Jacobi spectral problems.

Observe that the discrete modified string problem (3.7a)–(3.7c) has yet another representation. If we let

$$w = \left(-\sqrt{g}_1 \Pi_1, -\sqrt{h}_1 \Theta_1, \sqrt{g}_2 \Pi_2, \sqrt{h}_2 \Theta_2, \ldots, (-1)^K \sqrt{g}_K \Pi_K, (-1)^K \sqrt{h}_K \Theta_K\right)^T,$$

then (3.7a)–(3.7c) is equivalent to

$$Jw = \lambda w, \tag{6.1}$$

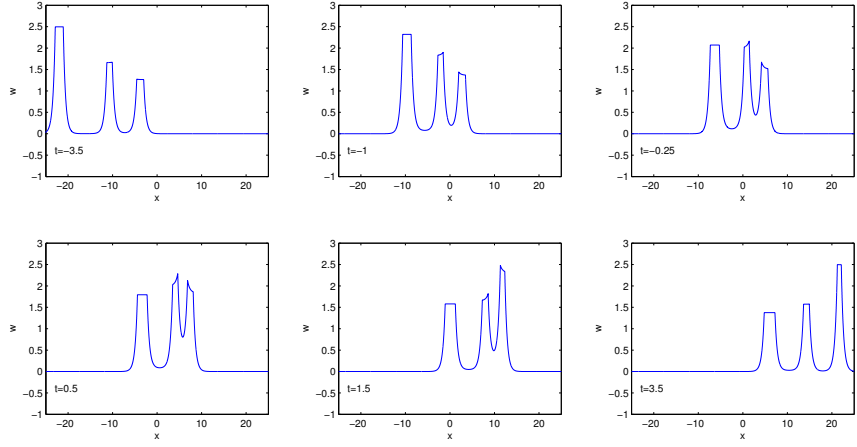

FIGURE 5. Graphs of $w(x,t) = u(x,t)v(x,t)$ for $K = 3$ at time $t = -3.5$, $-1, -0.25, 0.5, 1.5, 3$ in the case of $b_1(0) = 1$, $b_2(0) = 2$, $b_3(0) = 3$, $\zeta_1 = 0.4$, $\zeta_2 = 0.6$, $\zeta_2 = 0.8$, $m_1 = 15$, $n_2 = 1$, $m_3 = 0.14$, $n_4 = 40$, $m_5 = 0.06$, $n_6 = 100$.

where $J$ is a special Jacobi matrix

$$J = \begin{pmatrix} 0 & d_1 & 0 & \cdots & & 0 \\ d_1 & 0 & d_2 & 0 & & \\ 0 & d_2 & 0 & & & \\ \vdots & & & \ddots & & d_{2K-1} \\ 0 & & & & d_{2K-1} & 0 \end{pmatrix}_{2K \times 2K} \quad (6.2)$$

with

$$d_{2k-1} = \sqrt{\frac{1}{g_k h_k}}, \qquad d_{2k} = \sqrt{\frac{1}{g_{k+1} h_k}}. \quad (6.3)$$

Note that in this section we consider the positive roots in defining $d_k$. This kind of Jacobi spectral problem was studied in the work of Moser [34, 35]. We will briefly review the results pertaining to spectral properties of $J$. First, the spectrum of $J$ is real and simple. Additionally, if we define $\mathfrak{D}_k(\lambda)$ as the $k \times k$ principal minor of $\lambda I - J$ then

$$\mathfrak{D}_{k+1}(\lambda) = \lambda \mathfrak{D}_k(\lambda) - d_k^2 \mathfrak{D}_{k-1}(\lambda) \quad (6.4)$$

with $\mathfrak{D}_1 = \lambda$ and $\mathfrak{D}_0 = 1$. Moreover, the relation

$$\mathfrak{D}_k(-\lambda) = (-1)^k \mathfrak{D}_k(\lambda)$$

holds. Therefore, by noting that the characteristic polynomial equals $\mathfrak{D}_{2K}(\lambda)$ we conclude that if $\lambda$ is an eigenvalue of $J$ then so is $-\lambda$. We note that this provides another proof of Theorem 3.5.

**Remark 6.1.** Note that this kind of Jacobi matrix generates symmetric orthogonal polynomials in $\lambda$, namely those for which odd moments are zero. This fact is also the underlying reason why in the solution to the recurrence relations (3.7a), (3.7b), (3.7c), there are odd polynomials $\Theta_k$ and even polynomials $\Pi_k$.

The Weyl function associated to $J$ can be defined to be the element in the last row and last column of $(\lambda I - J)^{-1}$, i.e.,

$$W^*(\lambda) \stackrel{\text{def}}{=} \langle 2K|(\lambda I - J)^{-1}|2K\rangle = \frac{\mathfrak{D}_{2K-1}}{\mathfrak{D}_{2K}}. \tag{6.5}$$

The following properties of $W^*(\lambda)$ were proven in [34, 35] although they go back at least to T. Stieltjes [41].

**Theorem 6.2.** *[J. Moser]*

(1) $W^*(\lambda)$ *can be developed in a continued fraction*

$$W^*(\lambda) = \cfrac{1}{\lambda - \cfrac{d_{2K-1}^2}{\lambda - \cfrac{d_{2K-2}^2}{\lambda - \cdots - \cfrac{d_1^2}{\lambda}}}}. \tag{6.6}$$

(2) $W^*(\lambda)$ *admits the partial fraction decomposition*

$$W^*(\lambda) = \frac{1}{2}\sum_{j=1}^{K}\left(\frac{c_j}{\lambda - \lambda_j} + \frac{c_j}{\lambda + \lambda_j}\right) = \sum_{j=1}^{K}\frac{\lambda c_j}{\lambda^2 - \lambda_j^2}, \tag{6.7}$$

*with strictly positive residues $c_j$.*

(3)
$$\sum_{j=1}^{K} c_j = 1. \tag{6.8}$$

This concludes our review of pertinent work on the spectral properties of $J$. Our next goal is to establish a precise link between $J$ and the modified string problem.

**Theorem 6.3.** *The Weyl function $W(\lambda)$ of the modified string problem and the Weyl function $W^*(\lambda)$ of the associated Jacobi matrix satisfy*

$$W^*(\lambda) = -h_K \lambda W(\lambda).$$

*Proof.* In the notation of Section 3 and the definition of $w$ and $J$ from this section the first two equations (3.7a) and (3.7b) (without imposing boundary conditions) can be written, denoting again the vector $(0,0,\cdots,1) = |2K\rangle$,

$$\frac{(-1)^{K+1}\pi_{K+1}}{\sqrt{h_K}}|2K\rangle = (\lambda I - J)\,w,$$

which implies

$$(\lambda I - J)^{-1}|2K\rangle = \frac{(-1)^{K+1}\sqrt{h_K}}{\pi_{K+1}}w,$$

leading to the claim in view of the definition of $W(\lambda)$ given by (4.1). □

We have the following immediate consequence of Theorem 6.3:

**Corollary 6.4.** *For any $j = 1,\ldots,K$ the residues $c_j$ in $W^*(\lambda)$ and $b_j$ in $W(\lambda)$ satisfy*

$$c_j = h_K b_j. \tag{6.9}$$

*Proof.* The result follows from Theorem 6.3 upon comparing (6.7) with (4.3). □

Since $h_K > 0$ and $c_j > 0$, by Moser's result 6.2, we obtain the result announced in Theorem 4.3.

**Corollary 6.5.** *The residues $b_j$ in Theorem 4.3 are strictly positive.*

We end this subsection by formulating a correspondence between the modified string and the associated Jacobi matrix, as well the inhomogeneous string problems (4.8) and Jacobi spectral problems. First, we have the following theorem:

**Theorem 6.6.** *For a fixed $h_K$, the map*

$$(g_1, h_1, \ldots, g_{K-1}, h_{K-1}, g_K) \mapsto (d_1, \ldots, d_{2K-1})$$

*is a bijection between the modified string problem and the associated Jacobi matrix, which share the same spectral problem.*

*Proof.* The formulas (6.3) induces a map from the modified string to the associated Jacobi matrix.

Conversely, suppose that a Jacobi matrix (6.2) with positive $d_k$ is given. Then one can define the corresponding modified spectral problem (3.7a)–(3.7c) by letting

$$g_k = \frac{1}{h_K d_{2K-1}^2} \prod_{j=k}^{K-1} \left(\frac{d_{2j}}{d_{2j-1}}\right)^2, \qquad h_k = h_K \prod_{j=k}^{K-1} \left(\frac{d_{2j+1}}{d_{2j}}\right)^2.$$

□

Since inhomogeneous string problems of the form (4.8) are equivalent (see Remark 4.9) to modified string problems of the form (3.3)–(3.4), we have the following theorem:

**Theorem 6.7.** *Consider all discrete, finite, inhomogeneous string problems (4.8) whose spectral measures $d\mu$ are probability measures, i.e., $\int d\mu = 1$. Then these string problems are equivalent (see Remark 4.9) to spectral problems for Jacobi matrices of the type (6.2). In particular their respective Weyl functions satisfy*

$$\frac{W^*(\lambda)}{\lambda} = -W(\lambda) = \int \frac{d\mu(\zeta)}{\lambda^2 - \zeta},$$

*where $d\mu$ is the spectral measure of the string.*

*Proof.* The statement follows easily from $h_K = \frac{1}{\int d\mu}$ and Theorems 4.5, 6.3 and 6.6. □

**6.2. The Kac–van Moerbeke lattice as a Dirichlet/Neumann string.** Let us start with the string considered in Section 4.3. The mass density $\rho = \sum_{k=1}^{K} \rho_k \delta_{\xi_k}$ is mapped to the spectral density $d\mu = \sum_{k=1}^{K} b_k \delta_{\zeta_k}$. This map is a bijection as longs as $\rho_k > 0, b_k > 0, \zeta_k > 0, \zeta_j \neq \zeta_k, k \neq j$. We make an elementary observation.

**Lemma 6.8.** *For every string $S$ satisfying (4.8) with the spectral measure $d\mu$, there exists a unique string $S^*$ with the spectral measure $d\mu^*$, also satisfying (4.8), with the same spectrum and whose spectral measure is a probability measure. More explicitly,*

$$d\mu^* = \frac{d\mu}{\int d\mu}.$$

We will refer to such strings $S^*$ as *normalized* strings. As we proved earlier the 2-mCH equation implies the isospectral flow of the spectral measure given by (4.11) which we recall for the reader's convenience:

$$d\mu_t = e^{\frac{2t}{\zeta}} d\mu_0.$$

Thus we have the flow of normalized strings given by

$$d\mu_t^* = \frac{e^{\frac{2t}{\zeta}} d\mu_0}{\int e^{\frac{2t}{\zeta}} d\mu_0}.$$

By Theorem 6.7 normalized strings $S^*$ are in an isospectral correspondence with Jacobi matrices of type (6.2). Thus any flow of $d\mu^*$ induces a flow of Jacobi matrices of this type. To illustrate this point we will first consider a slightly different flow. Instead of the multiplier $e^{\frac{2t}{\zeta}}$ we will consider $e^{-2t\zeta}$, implying that the corresponding flow of normalized strings has spectral measures of the type $d\mu_t^* = \frac{e^{-2t\zeta} d\mu_0}{\int e^{-2t\zeta} d\mu_0}$. We will identify an explicit form of the corresponding isospectral deformations of Jacobi matrices. To this end we will briefly review the pertinent facts about such deformations.

For the Jacobi matrix (6.2), one can consider the isospectral deformations which satisfy the differential equation

$$\frac{dJ}{dt} = [A, J] \tag{6.10}$$

with the commutator $[A, J] = AJ - JA$. The matrix $A$ must be chosen appropriately to ensure that the commutator $[A, J]$ has zeros everywhere except on the two off-diagonals. The simplest possible choice [35] is the skew symmetric matrix

$$A = \begin{pmatrix} 0 & 0 & d_1 d_2 & 0 & \cdots & 0 \\ 0 & 0 & 0 & d_2 d_3 & \cdots & 0 \\ -d_1 d_2 & 0 & 0 & \ddots & \ddots & 0 \\ \vdots & \ddots & \ddots & \ddots & 0 & d_{2K-2} d_{2K-1} \\ & & \ddots & 0 & 0 & 0 \\ 0 & & & -d_{2K-2} d_{2K-1} & 0 & 0 \end{pmatrix}_{2K \times 2K}. \tag{6.11}$$

This matrix depends functionally on $J$, namely, if one decomposes $J^2$ into a lower triangular and a skew-symmetric part (the projection onto the skew-symmetric matrices taken along the strictly upper triangular matrices), then $A = J_{>0}^2 - (J_{>0}^2)^T$, where $M_{>0}$ refers to the strictly upper triangular part of a matrix $M$. With this choice of $A$, the differential system (6.10) takes the form

$$\dot{d}_k = d_k(d_{k+1}^2 - d_{k-1}^2), \qquad k = 1, \ldots, 2K-1 \tag{6.12}$$

with $d_0 = d_{2K} = 0$. Under the transformation

$$d_k = \frac{1}{2} e^{\frac{1}{2} v_k}, \tag{6.13}$$

one obtains

$$\dot{v}_k = \frac{1}{2}(e^{v_{k+1}} - e^{v_{k-1}}), \qquad k = 1, \ldots, 2K-1 \tag{6.14}$$

with the boundary condition $v_0 = v_{2K} = -\infty$. This is exactly the finite Kac–van Moerbeke lattice (or the finite Lotka-Volterra lattice, or the Langmuir lattice [44, 10]), which was derived as a discretization of the Korteweg–de Vries equation in the original work of M. Kac and P. van Moerbeke [25].

In [35], J. Moser described the solution to (6.12) in terms of rational functions of exponentials using the scattering method. In particular, he established (see also [25]) the formula for the evolution of the spectral measure. In terms of the residue $c_j$ the formula reads

$$c_j = \frac{c_j(0)e^{-2\lambda_j^2 t}}{\sum_{k=1}^{K} c_k(0)e^{-2\lambda_k^2 t}}. \tag{6.15}$$

Thus we obtain the next theorem.

**Theorem 6.9.** *The normalized string evolving according to $d\mu_t^* = \frac{e^{-2t\zeta} d\mu_0}{\int e^{-2t\zeta} d\mu_0}$ is mapped into the Kac–van Moerbeke finite lattice flow.*

Combining Theorem 4.5 and formulas (6.3) and (4.7), gives the following theorem:

**Theorem 6.10.** *The Kac–van Moerbeke lattice* (6.12) *admits the solution*

$$d_{2k-1} = \sqrt{\frac{H^1_{K+1-k} H^0_{K-k}}{H^0_{K+1-k} H^1_{K-k}}}, \qquad d_{2k} = \sqrt{\frac{H^1_{K-1-k} H^0_{K+1-k}}{H^1_{K-k} H^0_{K-k}}},$$

*where $H^l_k(t) = \det\bigl(A_{i+j+l}(t)\bigr)_{i,j=0}^{k-1}$. Here $A_k(t)$ is the kth moment of a string with spectral measure $d\mu_t = e^{-2t\zeta} d\mu_0$:*

$$A_k(t) = \int \zeta^k e^{-2t\zeta} d\mu_0 = \sum_{j=1}^{K} \lambda_j^{2k} b_j(t), \tag{6.16}$$

*with*

$$b_j(t) = b_j(0) e^{-2\lambda_j^2 t}$$

*and positive constants $\lambda_j$ and $b_j(0)$.*

**Remark 6.11.** The inverse formulas above have been known essentially since the times of Moser's paper [35] from which one sees that the solution may be expressed in terms of the rational functions of exponentials $e^{-2\lambda_j^2 t}$, $j = 1,\ldots,K$. The most general setup which results in this type of formulas is that of Yu. Berezanskii [5]. The only reason for presenting these formulas in our paper is that they come as a by-product of the natural connection between strings and lattices of Toda type. Observe also that the flow is actually linear on the spectral side of the string ($d\mu_t$ flow), not on the spectral side of the Jacobi matrix ($d\mu_t^*$).

**Corollary 6.12.** *The peakon flow* (2.12a)–(2.12b) *is mapped, under the correspondence described in Theorems 6.7 and 6.9, to the negative flow of the finite Kac–van Moerbeke lattice* (6.12).

We conclude by remarking that a similar relation between the peakon flow of the CH equation and the Toda lattice was described in [4].

## 7. Acknowledgements

The authors thank H. Lundmark for help with the presentation of results of this work.

## Appendix A. Lax pair for peakon ODEs

The purpose of this appendix is to give a rigorous interpretation for the Lax pair of the 2-mCH equation in the interlacing peakon solution case. The argument is similar to that in the Appendix B of [22]; in particular we use the same notation.

Let $\Omega_k$ denote the region $x_k(t) < x < x_{k+1}(t)$, where $x_k$ are smooth functions such that $-\infty = x_0(t) < x_1(t) < \cdots < x_N(t) < x_{N+1}(t) = +\infty$.

Let the function space $PC^\infty$ consist of all piecewise smooth functions $f(x,t)$ such that the restriction of $f$ to each region $\Omega_k$ is a smooth function $f_k(x,t)$ defined on an open neighbourhood of $\Omega_k$. Actually, for each fixed $t$, $f(x,t)$ defines a regular distribution $T_f(t)$ in the class $\mathscr{D}'(R)$ (for simplicity we will write $f$ instead of $T_f$). Note that the value of $f(x,t)$ on $x_k(t)$ does not need to be defined.

At every point $x_k$, let $f_x(x_k^-,t)$ and $f_x(x_k^+,t)$ denote the left and right limits of the function $f(x,t)$, and let us write

$$[f(x_k,t)] = f(x_k^+,t) - f(x_k^-,t), \qquad \langle f(x_k,t)\rangle = \frac{f(x_k^+,t)+f(x_k^-,t)}{2},$$

to denote the jump and the average, respectively. Denote by $f_x$ (or $f_t$) the ordinary (classical) partial derivative with respect to $x$ (or $t$), and by $\frac{\partial f_k}{\partial x}$ (or $\frac{\partial f_k}{\partial t}$) their restrictions to $\Omega_k$. If $D_x f$ denotes the distributional derivative with respect to $x$, then we have

$$D_x f = f_x + \sum_{k=1}^{N} [f(x_k)]\delta_{x_k}.$$

If we define $D_t f$ as the limit

$$D_t f(t) = \lim_{a\to 0}\frac{f(t+a)-f(t)}{a},$$

then we have

$$D_t f = f_t - \sum_{k=1}^{N} \dot{x}_k [f(x_k)]\delta_{x_k},$$

where $\dot{x}_k = dx_k/dt$.

Moreover, the following formulas also hold:

$$[fg] = \langle f\rangle[g] + [f]\langle g\rangle, \qquad \langle fg\rangle = \langle f\rangle\langle g\rangle + \frac{1}{4}[f][g] \tag{A.1}$$

$$\frac{d}{dt}[f(x_k)] = [f_x(x_k)]\dot{x}_k + [f_t(x_k)], \tag{A.2}$$

$$\frac{d}{dt}\langle f(x_k)\rangle = \langle f_x(x_k)\rangle\dot{x}_k + \langle f_t(x_k)\rangle, \tag{A.3}$$

for any $f,g \in PC^\infty$.

The interlacing peakon solution (5.1) belongs to the piecewise smooth class $PC^\infty$ with $N=2K$. The functions $\Psi_1$ and $\Psi_2$ in the Lax pair (2.2)–(2.3) are functions in $PC^\infty$. Actually, $u$, $u_x$ and $\Psi_1$ will be smooth functions in $x_{2k-1} < x < x_{2k+1}$, and $v$, $v_x$ and $\Psi_2$ will be smooth functions in $x_{2k} < x < x_{2k+2}$. Moreover, $u$ and $v$ are continuous throughout, while $u_x$ and $\Psi_1$ have a jump at each $x_{2k-1}$, and $v_x$ and $\Psi_2$ have a jump at each $x_{2k}$.

Let $\Psi = (\Psi_1, \Psi_2)$, and consider the overdetermined system

$$D_x \Psi = \frac{1}{2}\hat{L}\Psi, \qquad D_t \Psi = \frac{1}{2}\hat{A}\Psi,$$

where

$$\hat{L} = L + 2\lambda\left(\sum_{k=1}^{K} m_{2k-1}\delta_{x_{2k-1}}\right)N_1 - 2\lambda\left(\sum_{k=1}^{K} n_{2k}\delta_{x_{2k}}\right)N_2, \tag{A.4}$$

and

$$\hat{A} = A - 2\lambda\left(\sum_{k=1}^{K} m_{2k-1}Q(x_{2k-1})\delta_{x_{2k-1}}\right)N_1 + 2\lambda\left(\sum_{k=1}^{K} n_{2k}Q(x_{2k})\delta_{x_{2k}}\right)N_2, \tag{A.5}$$

with

$$L = \begin{pmatrix} -1 & 0 \\ 0 & 1 \end{pmatrix}, \qquad N_1 = \begin{pmatrix} 0 & 1 \\ 0 & 0 \end{pmatrix}, \qquad N_2 = \begin{pmatrix} 0 & 0 \\ 1 & 0 \end{pmatrix}, \tag{A.6}$$

$$A = \begin{pmatrix} 4\lambda^{-2} + Q & -2\lambda^{-1}(u - u_x) \\ 2\lambda^{-1}(v + v_x) & -Q \end{pmatrix}, \tag{A.7}$$

and $Q = (u - u_x)(v + v_x)$. Note that (A.4) involves multiplying $N_1 \Psi = (\Psi_2, 0)$ by $\delta_{x_{2k-1}}$ and $N_2 \Psi = (0, \Psi_1)$ by $\delta_{x_{2k}}$ which is legitimate in view of the interlacing condition. In fact, this would work for any measures $m$ and $n$ with non-overlapping singular supports. Thus the $x$-Lax equation is well defined as a distributional equation. This is not the case with the $t$-Lax equation. Indeed, for (A.5) to be defined as a distributional equation, $u_x N_1 \Psi = (u_x \Psi_2, 0)$ needs to be a multiplier of $\delta_{x_{2k-1}}$; likewise, $v_x N_2 \Psi = (0, v_x \Psi_1)$ needs to be a multiplier of $\delta_{x_{2k}}$. Thus the values of $u_x(x_{2k-1})$ and $v_x(x_{2k})$ need to be assigned, since $\Psi_1(x_{2k})$ and $\Psi_2(x_{2k-1})$ are already assigned by continuity. The following statement indicates that the Lax pair is well defined and compatible as two distributional equations when $u_x(x_{2k-1})$ and $v_x(x_{2k})$ are all defined to be the averages of their respective left and right-hand limits. Again, it is not necessary to assume the interlacing condition. Instead, the non-overlapping support of $m$ and $n$ suffices.

**Theorem A.1.** *Let $m$ and $n$ be two discrete measures associated to $u$ and $v$ given by* (2.6) *and* (2.7). *Suppose that $u_x(x_k)$ and $v_x(x_k)$ are assigned values*

$$u_x(x_k) = \langle u_x(x_k) \rangle, \qquad v_x(x_k) = \langle v_x(x_k) \rangle,$$

*and, subsequently, $Q(x_k) = \langle Q(x_k) \rangle$. Then the compatibility condition $D_t D_x \Psi = D_x D_t \Psi$ of the Lax pair* (A.4)–(A.5) *reads*

$$\dot{m}_k = 0, \qquad \dot{n}_k = 0, \qquad \dot{x}_k = \langle Q(x_k) \rangle,$$

*which is equivalent to* (2.9) *and* (2.10).

*Proof.* The proof proceeds in a similar way to Theorem B.1 in [22]. The details are omitted here. □

MULTIPEAKONS OF A 2-MCH EQUATION AND THE FINITE KVM LATTICE 27

LSEC, Institute of Computational Mathematics and Scientific Engineering Computing, AMSS, Chinese Academy of Sciences, P.O.Box 2719, Beijing 100190, PR China, and the Department of Mathematics and Statistics, University of Saskatchewan, 106 Wiggins Road, Saskatoon, Saskatchewan, S7N 5E6, Canada.
  *E-mail address*: `changxk@lsec.cc.ac.cn`

LSEC, Institute of Computational Mathematics and Scientific Engineering Computing, AMSS, Chinese Academy of Sciences, P.O.Box 2719, Beijing 100190, PR China.
  *E-mail address*: `hxb@lsec.cc.ac.cn`

Department of Mathematics and Statistics, University of Saskatchewan, 106 Wiggins Road, Saskatoon, Saskatchewan, S7N 5E6, Canada.
  *E-mail address*: `szmigiel@math.usask.ca`